\documentclass[12pt]{article}
\usepackage{amsmath,caption}
\usepackage[top=1.35in, bottom=1.35in, left=1.35in, right=1.35in]{geometry}

\DeclareCaptionStyle{italic}[justification=centering]{labelfont={bf},textfont={it},labelsep=colon}
\usepackage{graphicx,psfrag,epsf}
\usepackage{enumerate}
\usepackage{natbib}
\usepackage{url} 
\usepackage{booktabs, subfig, bm, paralist,mathpazo,tikz,todonotes,longtable,microtype} 
\usepackage[pdftex,colorlinks=true]{hyperref}
\definecolor{darkblue}{rgb}{0,0,.6}
\hypersetup{citecolor=darkblue,linkcolor=darkblue,urlcolor=darkblue}

\newcommand{\blind}{0}

\graphicspath{{plots/}}
\DeclareMathOperator*{\argmin}{\arg\!\min}
\newsavebox\CBox
\def\textBF#1{\sbox\CBox{#1}\resizebox{\wd\CBox}{\ht\CBox}{\textbf{#1}}}

\definecolor{a0}{rgb}{0.0, 0.5, 0.0}
\definecolor{bistre}{rgb}{0.24, 0.17, 0.12}
\definecolor{amethyst}{rgb}{0.6, 0.4, 0.8}
\definecolor{blue-violet}{rgb}{0.54, 0.17, 0.89}
\definecolor{Rcolor}{RGB}{150,160,190}
\definecolor{blush}{rgb}{0.87, 0.36, 0.51}
\definecolor{brightturquoise}{rgb}{0.03, 0.91, 0.87}
\definecolor{burntorange}{rgb}{0.8, 0.33, 0.0}
\begin{document}

\def\spacingset#1{\renewcommand{\baselinestretch}%
{#1}\small\normalsize} \spacingset{1}

\if0\blind
{
  \title{\bf Mortality and life expectancy forecasting for a group of populations in developed countries: \hbox{A robust multilevel functional data method}}
  \author{Han Lin Shang\thanks{Postal address: Research School of Finance, Actuarial Studies and Statistics, Australian National University, Canberra ACT 2601, Australia; Telephone number: +61-2-6125 0535; Fax number: +61-2-6125 0087; E-mail: hanlin.shang@anu.edu.au.}}
  \maketitle
} \fi

\if1\blind
{
  \bigskip
  \bigskip
  \bigskip
  \begin{center}
    {\LARGE\bf Title}
\end{center}
  \medskip
} \fi

\bigskip
\begin{abstract}
A robust multilevel functional data method is proposed to forecast age-specific mortality rate and life expectancy for two or more populations in developed countries with high-quality vital registration systems. It uses a robust multilevel functional principal component analysis  of aggregate and population-specific data to extract the common trend and population-specific residual trend among populations. This method is applied to age- and sex-specific mortality rate and life expectancy for the United Kingdom from 1922 to 2011, and its forecast accuracy is then further compared with standard multilevel functional data method. For forecasting both age-specific mortality and life expectancy, the robust multilevel functional data method produces more accurate point and interval forecasts than the standard multilevel functional data method in the presence of outliers.
\end{abstract}

\vfill

\newpage
\spacingset{1.45}



\section{Introduction}\label{Shang:sec:1}

Many statistical methods have been proposed for forecasting age-specific mortality rates \citep[see][for reviews]{Booth06,BT08,SBH11,BT14}. Of these, a significant milestone in demographic forecasting was the work by \citet{LC92}. They applied a principal component method to age-specific mortality rates and extracted a single time-varying index of the level of mortality rates, from which the forecasts are obtained by a random-walk with drift. The method has since been extended and modified. For example, \citet{RH03} proposed the age-period-cohort Lee-Carter method; \citet{HU07} proposed a functional data model that utilizes nonparametric smoothing and high-order principal components; \citet{GK08} and \cite{WSB+15} considered Bayesian techniques for Lee-Carter model estimation and forecasting; and \citet{LLG13} extended the Lee-Carter method to model the rotation of age patterns for long-term projections.

These works mainly focused on forecasting age-specific mortality for a single population, or several populations individually. However, joint modeling mortality for two or more populations simultaneously is paramount, as it allows one to model the correlations among two or more populations, distinguish between long-term and short-term effects in the mortality evolution, and explore the additional information contained in the experience of other populations to further improve point and interval forecast accuracy. These populations can be grouped by sex, state, ethnic group, socioeconomic status and other attributes \citep[e.g.,][]{LL05, ARG+11, RLS+12, RCG+13, Li13,RLG14, SLK+15}.

As an extension of \citet{LL05}, we consider a robust multilevel functional data model described in Section~\ref{Shang:sec:2} by extending the work of \citet{DCC+09}, \citet{CSD09}, \citet{CG10}, \citet{GCC+10} and \cite{Shang16}. The objective of the multilevel functional data method is to model multiple sets of functions that may be correlated among groups. In this paper, we propose a robust version of this technique to forecast age-specific mortality and life expectancy for a group of populations. We found that the robust multilevel functional data model captures the correlation among populations, models the forecast uncertainty through Bayesian paradigm, and is adequate for use within a probabilistic population modeling framework \citep{RLS+12}. Similar to the work of \citet{LL05}, \citet{Lee06} and \citet{DDG+06}, the robust multilevel functional data model captures the common trend and the population-specific residual trend. It produces forecasts that are more accurate than the ones from the standard multilevel functional data method, in the presence of outliers. Illustrated by the age- and sex-specific mortality rates for the United Kingdom, we study and compare the performance of the standard and robust multilevel functional data methods in Section~\ref{Shang:sec:3}. In Section~\ref{Shang:sec:4}, we provide some concluding remarks.

\section{A robust multilevel functional data method}\label{Shang:sec:2}

We present this method in the context of forecasting female and male age-specific mortality in a country, although the method can easily be generalized to any number of sub-populations. Let $y_t^j(x_i)$ be the log central mortality of the $j^{\text{th}}$ population for year $t=1,2,\dots,n$ at observed ages $x_1,x_2,\dots,x_p$ where $x$ is a continuous variable and $p$ is the number of ages.

Since we consider forecasting age-specific mortality from a functional data analytic viewpoint, each function should be smooth and continuous. A nonparametric smoothing technique is thus implemented to construct a time series of functions $\left\{f_1^j(x),f_2^j(x),\dots,f_n^j(x)\right\}$. That is
\begin{equation}
y_t^j(x_i) = f_t^j(x_i)+\delta_t^j(x_i)\varepsilon_{t,i}^j, \label{Shang:eq:1}
\end{equation}
where $x_i$ represents the centre of each age or age group for $i=1,2,\dots,p$, $\varepsilon_{t,i}^j$ is an independent and identically distributed (iid) standard normal random variable, $\delta_t^j(x_i)$ captures different variances for different ages. Together, $\delta_t^j(x_i)\varepsilon_{t,i}^j$ represents the smoothing error (also known as measurement error).

Let $m_t^j(x_i) = \exp\{y_t^j(x_i)\}$ be the observed central mortality for ages $x_i$ at year $t$ and let $N_t(x_i)$ be the total mid-year population of age $x_i$ in year $t$. The observed mortality rate approximately follows a binomial distribution with estimated variance
\begin{equation*}
\text{Var}\left[m_t^j(x_i)\right]\approx \frac{m_t^j(x_i)\times \left(1-m_t^j\left(x_i\right)\right)}{N_t(x_i)}.
\end{equation*}
Via Taylor's series expansion, the estimated variance associated with the log mortality rate is given by
\begin{equation*}
\left(\widehat{\delta}_t^j\right)^2(x_i) =\text{Var}\left[\ln\left(m_t^j(x_i)\right)\right]\approx\frac{1-m_t^j(x_i)}{m_t^j(x_i)\times N_t(x_i)}.
\end{equation*}
Since $m_t^j(x_i)$ is often quite small, $\left(\delta_t^j\right)^2(x_i)$ can also be approximated by a Poisson distribution with estimated variance
\begin{equation*}
\left(\widehat{\delta}_t^j\right)^2(x_i)\approx \frac{1}{m_t^j(x_i)\times N_t(x_i)}.
\end{equation*}

Let the weights be the inverse variances $w_t(x_i) = 1/\left(\widehat{\delta}_t^j\right)^2(x_i)$, the log mortality rates are smoothed by using weighted penalized regression spline with a partial monotonic constraint for ages above 65 \citep{HU07}. The penalized regression spline can be written as:
 \begin{equation*}
 \widehat{f}_t(x_i) = \argmin_{\theta_t(x_i)}\sum_{i=1}^Mw_t(x_i)\left|y_t(x_i)-\theta_t(x_i)\right|+\alpha\sum^{M-1}_{i=1}\left|\theta_t^{'}(x_{i+1})-\theta_t^{'}(x_i)\right|, 
 \end{equation*}
where $i$ represents different ages (grid points) in a total of $M$ grid points, $\alpha$ represents a smoothing parameter, and $^{'}$ symbolizes the first derivative of a function. While the $L_1$ loss function and the $L_1$ roughness penalty are employed to obtain robust estimates, the monotonic increasing constraint helps to reduce the noise from estimation of high ages \citep[see also][]{HN99}. 

In the multilevel functional data method, we first apply~\eqref{Shang:eq:1} to smooth different sets of functions from different populations that may be correlated. In the case of two populations, the essence is to decompose curves among two populations into an average of total mortality, denoted by $\mu(x)$, a sex-specific deviation from the averaged total mortality, denoted by $\eta^j(x)$, a common trend that is shared by all populations, denoted by $R_t(x)$, a sex-specific trend that is specific to $j^{\text{th}}$ population, denoted by $U_t^j(x)$, and model error $e_t^j(x)$ with finite variance $(\sigma^2)^j$. The common trend is obtained by projecting a functional time series onto the eigenvectors of covariance operators of the aggregate and centered stochastic process. The sex-specific trend is then obtained by projecting the residual functions from the first eigen decomposition, onto the eigenvectors of covariance operators of the sex-specific and centered stochastic process. To express our idea mathematically, the smoothed log mortality rates at year $t$ can be written as:
\begin{equation}
f_t^j(x) = \mu(x) + \eta^j(x) + R_t(x) + U_t^j(x) + e_t^j(x),\qquad x\in \mathcal{I},\label{Shang:eq:4}
\end{equation}
for each $t$, where $\mathcal{I}$ represents a function support.

Since $R(x)$ and $U^j(x)$ are unknown in practice, they can be approximated by a set of realizations $\bm{R}(x) = \left\{R_1(x),R_2(x),\dots,R_n(x)\right\}$ and $\bm{U}^j(x) = \left\{U_1^j(x),U_2^j(x),\dots,U_n^j(x)\right\}$. Thus, the sample mean function of total mortality and sex-specific mortality can be expressed as:
\begin{eqnarray}
\widehat{\mu}(x) &= \frac{1}{n}\sum^n_{t=1}f_t^{\text{T}}(x) \label{Shang:eq:5}\\
\widehat{\mu}^j(x) &= \frac{1}{n}\sum^n_{t=1}f_t^j(x) \label{Shang:eq:6}\\
\widehat{\eta}^j(x) &= \widehat{\mu}^j(x) - \widehat{\mu}(x)\label{Shang:eq:7}
\end{eqnarray}
where $\{f_1^{\text{T}}(x),f_2^{\text{T}}(x),\dots,f_n^{\text{T}}(x)\}$ represents a set of smooth functions for the age-specific total mortality; $\widehat{\mu}(x)$ represents the simple average of smoothed total mortality; whereas $\widehat{\mu}^j(x)$ represents the simple average of smoothed male or female mortality; and $\widehat{\eta}^j(x)$ represents the difference between the mean of total mortality and the mean of sex-specific mortality.

Then, we consider a two-step algorithm by combining a robust functional principal component analysis and binary weighting. This can be described as:
\begin{enumerate}
\item[(1)] Use a robust principal component analysis, such as RAPCA \citep{HRV02} or ROBPCA \citep{HRB05}, to obtain initial (highly robust) values for $\left\{\widehat{\beta}_{t,k}\right\}$ and $\left\{\widehat{\phi}_k(x)\right\}$ for $t=1,\dots,n$ and $k=1,\dots,K$.
\item[(2)] Define the integrated squared error for year $t$ as:
\begin{equation*}
v_t = \int_{x\in\mathcal{I}} \Big[f_t(x) - \sum^K_{k=1}\widehat{\beta}_{t,k}\widehat{\phi}_k(x)\Big]^2dx.
\end{equation*}
\end{enumerate}
It identifies those outlying years that have higher values of $v_t$. We then assign weights $w_t=1$ if $v_t<s+\lambda\sqrt{s}$ and $w_t=0$ otherwise, where $s$ is the median of $\{v_1, v_2, \dots,v_n\}$ and $\lambda>0$ is a tuning parameter to control the efficiency of this robust algorithm. When $\lambda=3$, it represents $\Phi(3/\sqrt{2}) = 98.3\%$ efficiency, where the number of outliers is 1.7\% of total number of observations. When $\lambda\rightarrow \infty$, there is no outlier in the data; when $\lambda \rightarrow 0$, all observations are identified as outliers. For $\lambda>0$, this algorithm retains the optimal breakdown point of 0.5.

Having obtained a set of robust basis functions, the common and sex-specific trends can be estimated by:
\begin{eqnarray}
\widehat{R}_t(x) \approx \sum^K_{k=1}\widehat{\beta}_{t,k}\widehat{\phi}_k(x), \notag\\
\widehat{U}_t^j(x) \approx \sum^L_{l=1}\widehat{\gamma}_{t,l}^j\widehat{\psi}_l^j(x),\label{Shang:eq:9}
\end{eqnarray}
where $\big\{\widehat{\bm{\beta}}_k=\big(\widehat{\beta}_{1,k},\widehat{\beta}_{2,k},\dots,\widehat{\beta}_{n,k}\big);k=1,\dots,K\big\}$ represents the $k^{\text{th}}$ sample principal component scores of $R(x)$; $\bm{\Phi} = \big[\widehat{\phi}_1(x),\widehat{\phi}_2(x),\dots,\widehat{\phi}_K(x)\big]$ are the corresponding orthogonal sample eigenfunctions in a square integrable function space. Similarly, $\{\widehat{\bm{\gamma}}_l^j = (\widehat{\gamma}_{1,l}^j,\widehat{\gamma}_{2,l}^j,\dots,\widehat{\gamma}_{n,l}^j); l=1,\dots,L\}$ represents the $l^{\text{th}}$ sample principal component scores of $U^j(x)$, and $\bm{\Psi} = [\widehat{\psi}_1^j(x),\widehat{\psi}_2^j(x),\dots,\widehat{\psi}_L^j(x)]$ are the corresponding orthogonal sample eigenfunctions. Since two stochastic processes $R(x)$ and $U^j(x)$ are uncorrelated, $\widehat{\bm{\beta}}_k$ are uncorrelated with $\widehat{\bm{\gamma}}_l^j$.

It is important to select optimal $K$ and $L$, and three common approaches are leave-one-out or leave-more-out cross validation \citep{RS91}, Akaike information criterion \citep{YMW05} and explained variance \citep{CG10,Chiou12}. We use a cumulative percentage of total variation to determine $K$ and $L$. The optimal numbers of $K$ and $L$ are determined by:
\begin{align*}
K&=\argmin_{K:K\geq 1}\left\{\sum^K_{k=1}\lambda_k\Big/\sum^{\infty}_{k=1}\lambda_k\geq P\right\}, \\
L&=\argmin_{L:L\geq 1}\left\{\sum^L_{l=1}\lambda_l^j\Big/\sum^{\infty}_{l=1}\lambda_l^j\geq P\right\}, \qquad \text{for each \ $j$.}
\end{align*}  
Following \citet{CG10} and \citet{Chiou12}, we chose $P=0.9$.

An important parameter in the multilevel functional data method is the proportion of variability explained by aggregate data, which is the variance explained by the within-cluster variability \citep{DCC+09}. A possible measure of within-cluster variability is given by:
\begin{equation}
\frac{\sum^{\infty}_{k=1}\lambda_k}{\sum^{\infty}_{k=1}\lambda_k+\sum^{\infty}_{l=1}\lambda^j_l}=\frac{\int_{\mathcal{I}}\text{var}\{R(x)\}dx}{\int_{\mathcal{I}}\text{var}\{R(x)\}dx+\int_{\mathcal{I}}\text{var}\{U^j(x)\}dx}. \label{Shang:eq:2}
\end{equation}
When the common factor can explain the main mode of total variability, the value of within-cluster variability is close to 1.

Substituting equations~\eqref{Shang:eq:5}--~\eqref{Shang:eq:9} into equations~\eqref{Shang:eq:4}--~\eqref{Shang:eq:1}, we obtain
\begin{equation*}
y_t^j(x) = \widehat{\mu}(x) + \widehat{\eta}^j(x) + \sum^K_{k=1}\widehat{\beta}_{t,k}\widehat{\phi}_k(x) + \sum^L_{l=1}\widehat{\gamma}_{t,l}^j\widehat{\psi}_l^j(x) + e_t^j(x) + \delta_t^j(x)\varepsilon_t^j,\label{Shang:eq:10}
\end{equation*}
where $\widehat{\beta}_{t,k}\sim \text{N}\big(0,\widehat{\lambda}_k\big)$, $\widehat{\gamma}_{t,l}^j\sim \text{N}\big(0,\widehat{\lambda}_l^j\big)$, $e_t^j(x)\sim\text{N}\big(0,(\widehat{\sigma}^2)^j\big)$ and $\widehat{\lambda}_k$ denotes the $k^{\text{th}}$ eigenvalue of estimated covariance operator associated with the common trend, and $\widehat{\lambda}_l^j$ represents the $l^{\text{th}}$ eigenvalue of estimated covariance operator associated with the sex-specific residual trend.

Conditioning on the estimated functional principal components $\bm{\Phi}$, $\bm{\Psi}$ and continuous functions $\bm{y}^j = [y_1^j(x),y_2^j(x),\dots,y_n^j(x)]$, the $h$-step-ahead point forecasts of $y_{n+h}^j(x)$ are given by:
\begin{align*}
\widehat{y}_{n+h|n}^j(x) &= \text{E}\left[y_{n+h}(x)|\bm{\Phi},\bm{\Psi},\bm{y}^j\right] \\
&=\widehat{\mu}(x) + \widehat{\eta}^j(x) + \sum^K_{k=1}\widehat{\beta}_{n+h|n,k}\widehat{\phi}_k(x) + \sum^L_{l=1}\widehat{\gamma}_{n+h|n,l}^j\widehat{\psi}^j_l(x),
\end{align*}
which $\widehat{\beta}_{n+h|n,k}$ and $\widehat{\gamma}_{n+h|n,l}^j$ are forecast univariate principal component scores, obtained from a univariate time series forecasting method, such as random walk with drift (rwf), exponential smoothing (ets), and autoregressive integrated moving average (ARIMA$(p,d,q)$) in which its optimal orders $p, d, q$ are determined automatically using an information criterion, such as corrected Akaike information criterion.

If $\{\widehat{\gamma}_{n+h|n,l}^{1}-\widehat{\gamma}_{n+h|n,l}^2; l=1,\dots,L\}$ has a trending long-term mean, the multilevel functional data method does not produce convergent forecasts. However, if the common mean function and common trend capture the long-term effect, the multilevel functional data method produces convergent forecasts, where the forecasts of residual trends would be flat.

To measure forecast uncertainty, the interval forecasts of $y_{n+h}^j(x)$ can be obtained through a Bayesian paradigm equipped with Markov chain Monte Carlo (MCMC). \cite{DCC+09} present a derivation of posterior of principal component scores, where MCMC is used to estimate all variance parameters and to draw samples from the posterior of principal component scores. The bootstrapped forecasts are given by:
\begin{align}
\widehat{y}_{n+h|n}^{b,j}(x) = & \widehat{\mu}(x) + \widehat{\eta}^j(x) + \sum^K_{k=1}\widehat{\beta}_{n+h|n,k}^b\widehat{\phi}_k(x) + \sum^L_{l=1}\widehat{\gamma}_{n+h|n,l}^{b,j}\widehat{\psi}_l^j(x) + \notag\\
& \widehat{e}_{n+h}^{b,j}(x) + \widehat{\delta}_{n+h}^{b,j}(x)\varepsilon_{n+h}^j,\label{Shang:eq:multilevel_interval}
\end{align}
for $b=1,\dots,B$. As previously studied by \citet[][supplementary materials]{DCC+09}, we first simulate $\{\widehat{\beta}_{1,k}^b,\dots,\widehat{\beta}_{n,k}^b\}$ drawn from its posterior, and then obtain $\widehat{\beta}_{n+h|n,k}^b$ using a univariate time series forecasting method for each simulated sample; similarly, we first simulate $\{\widehat{\gamma}_{1,l}^{b,j},\dots,\widehat{\gamma}_{n,l}^{b,j}\}$ drawn from its posterior, and then obtain $\widehat{\gamma}_{n+h|n,l}^{b,j}$ for each bootstrap sample; $\widehat{e}_{n+h}^{b,j}(x)$ is drawn from $N(0,(\widehat{\sigma}^2)^{b,j})$, where $(\widehat{\sigma}^2)^{b,j}$ is estimated at each iteration of MCMC. Since we pre-smooth functional data, we must add the smoothing error $\widehat{\delta}_{n+h}^{b,j}(x)$ as another source of randomness and $\varepsilon_{n+h}^j$ is drawn from $N(0,1)$ and $B=1000$ represents the number of MCMC draws. The prediction interval is constructed from the percentiles of the bootstrapped mortality forecasts. The interval forecasts of life expectancy are obtained from the forecast age-specific mortality using the life table method \citep{PHG01}.

\section{Application to the UK's age- and sex-specific mortality}\label{Shang:sec:3}

\begin{figure}[!t]
\centering
\subfloat[Log mortality rates]
{\includegraphics[width=8.42cm]{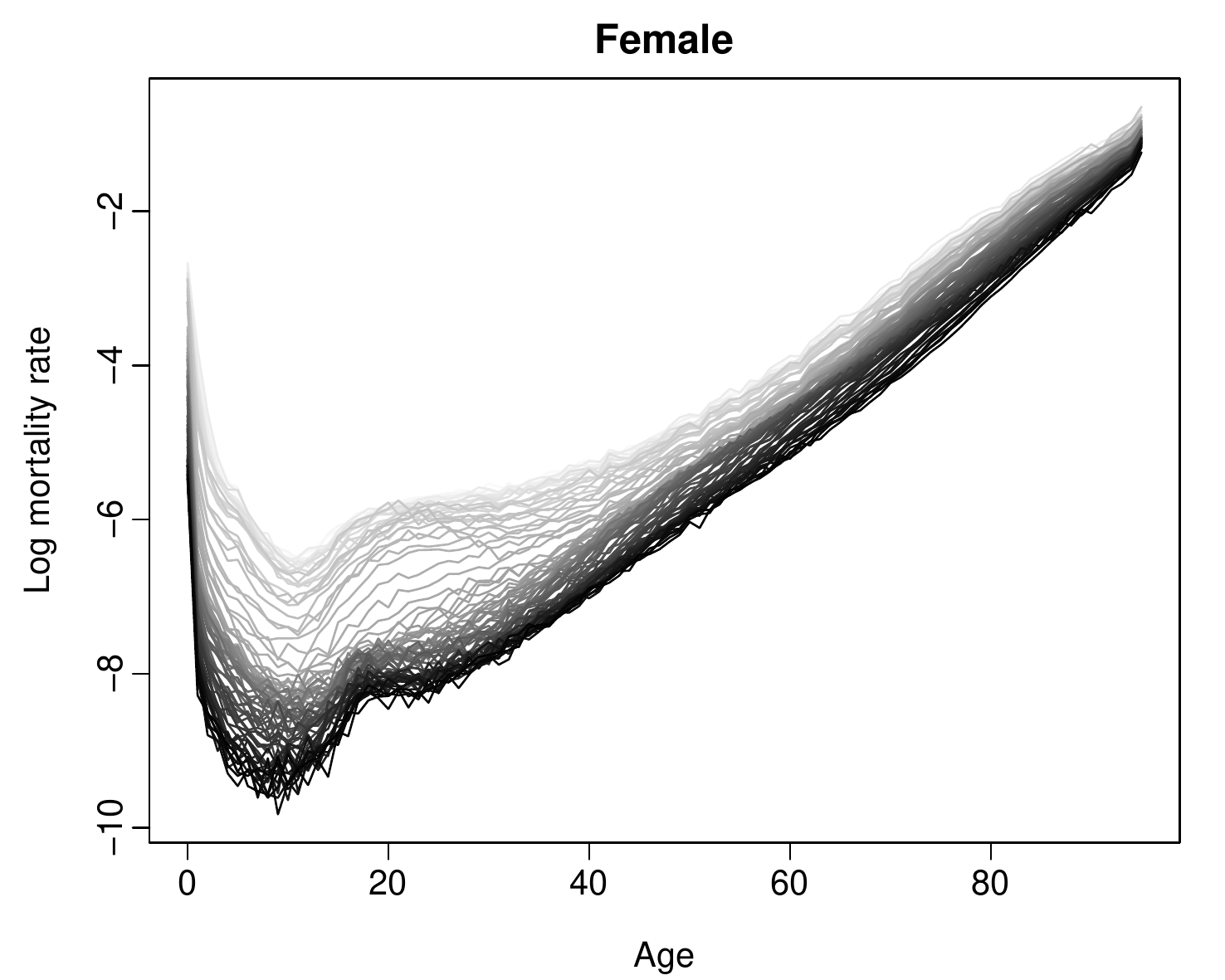}\label{Shang:fig:1a}}
\quad
\subfloat[Log mortality rates]
{\includegraphics[width=8.42cm]{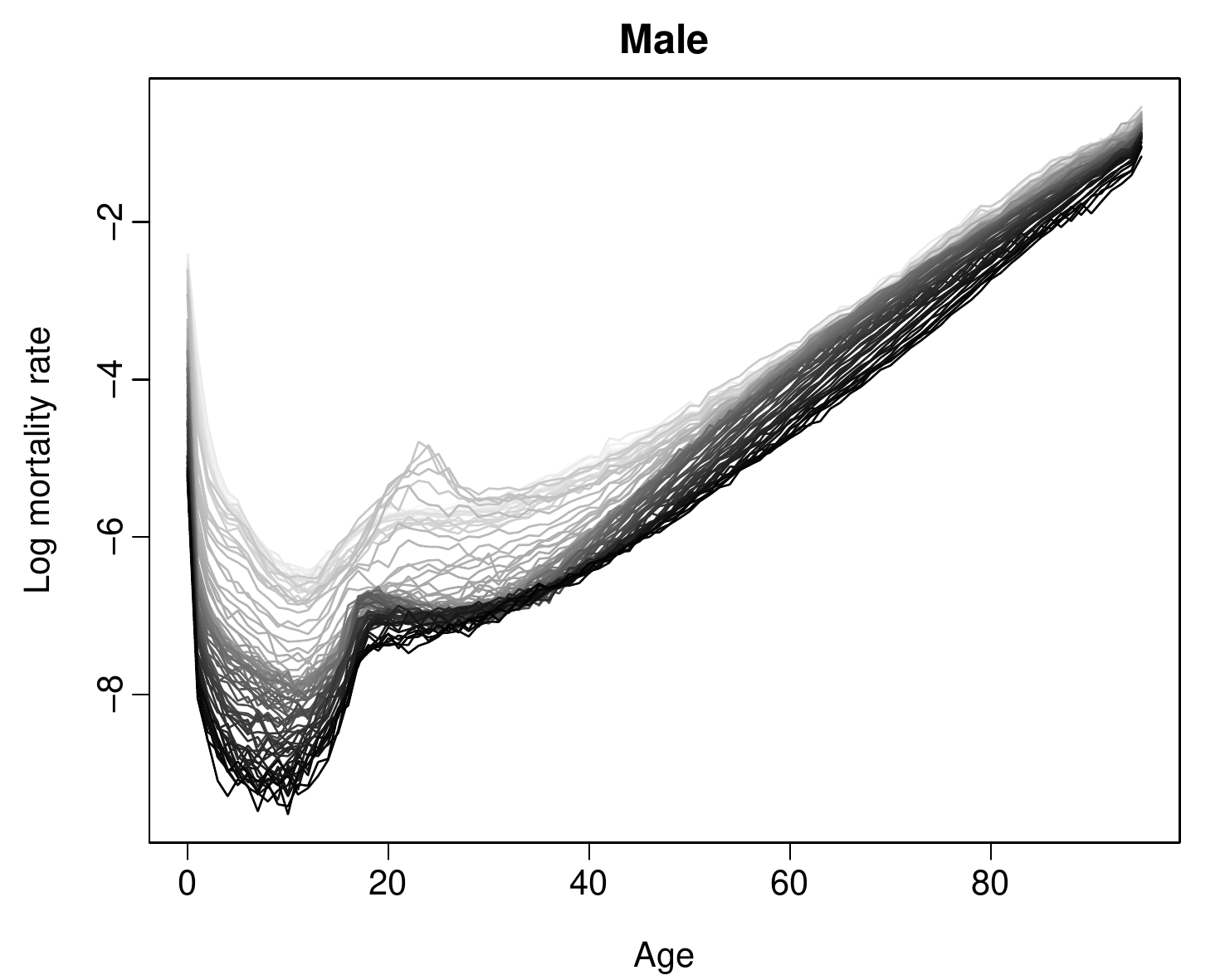}\label{Shang:fig:1b}}
\\
\subfloat[Smoothed log mortality rates]
{\includegraphics[width=8.42cm]{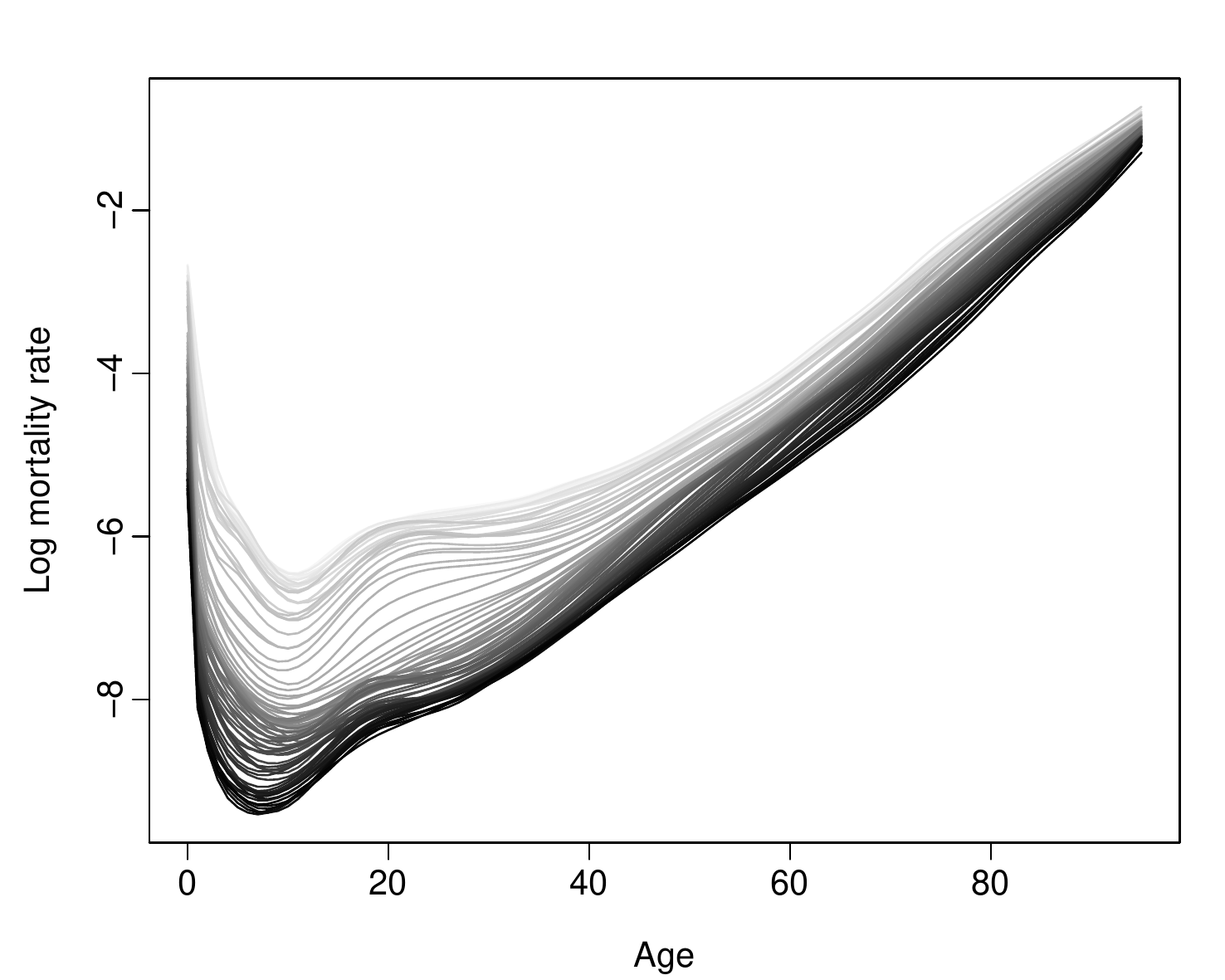}\label{Shang:fig:1c}}
\quad
\subfloat[Smoothed log mortality rates]
{\includegraphics[width=8.42cm]{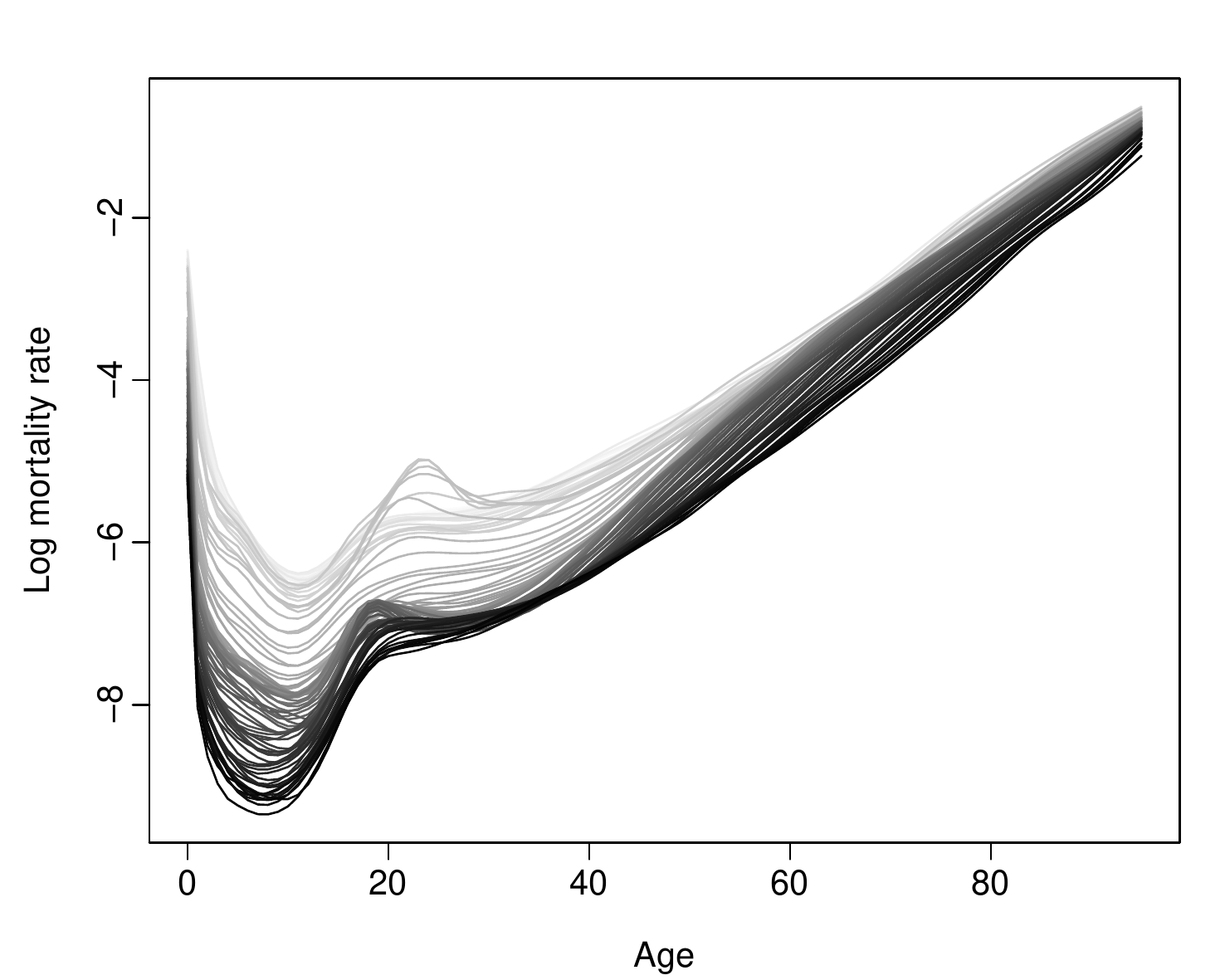}\label{Shang:fig:1d}}
\caption{Observed and smoothed age-specific female and male log mortality rates for the United Kingdom. Data from the distant past are shown in light gray, and the most recent data are shown in dark gray.}\label{Shang:fig:1}
\end{figure}

Age- and sex-specific mortality rates for the United Kingdom between 1922 and 2009 are available from the \citet{HMD13}. For each sex in a given calendar year, the mortality rates obtained by the ratio between ``number of deaths" and ``exposure to risk", are organized in a matrix by age and calendar year. By analyzing the changes in mortality as a function of age $x$ and year $t$, it can be seen that age-specific mortality rates have shown a gradual decline over years. In Figs~\ref{Shang:fig:1a} and~\ref{Shang:fig:1b}, we present functional time series plots of female and male log mortality rates. By using a weighted penalized regression spline, the smoothed female and male log mortality rates are obtained in Figs~\ref{Shang:fig:1c} and~\ref{Shang:fig:1d}. 

\begin{figure}[!htbp]
  \centering
  \includegraphics[width=\textwidth]{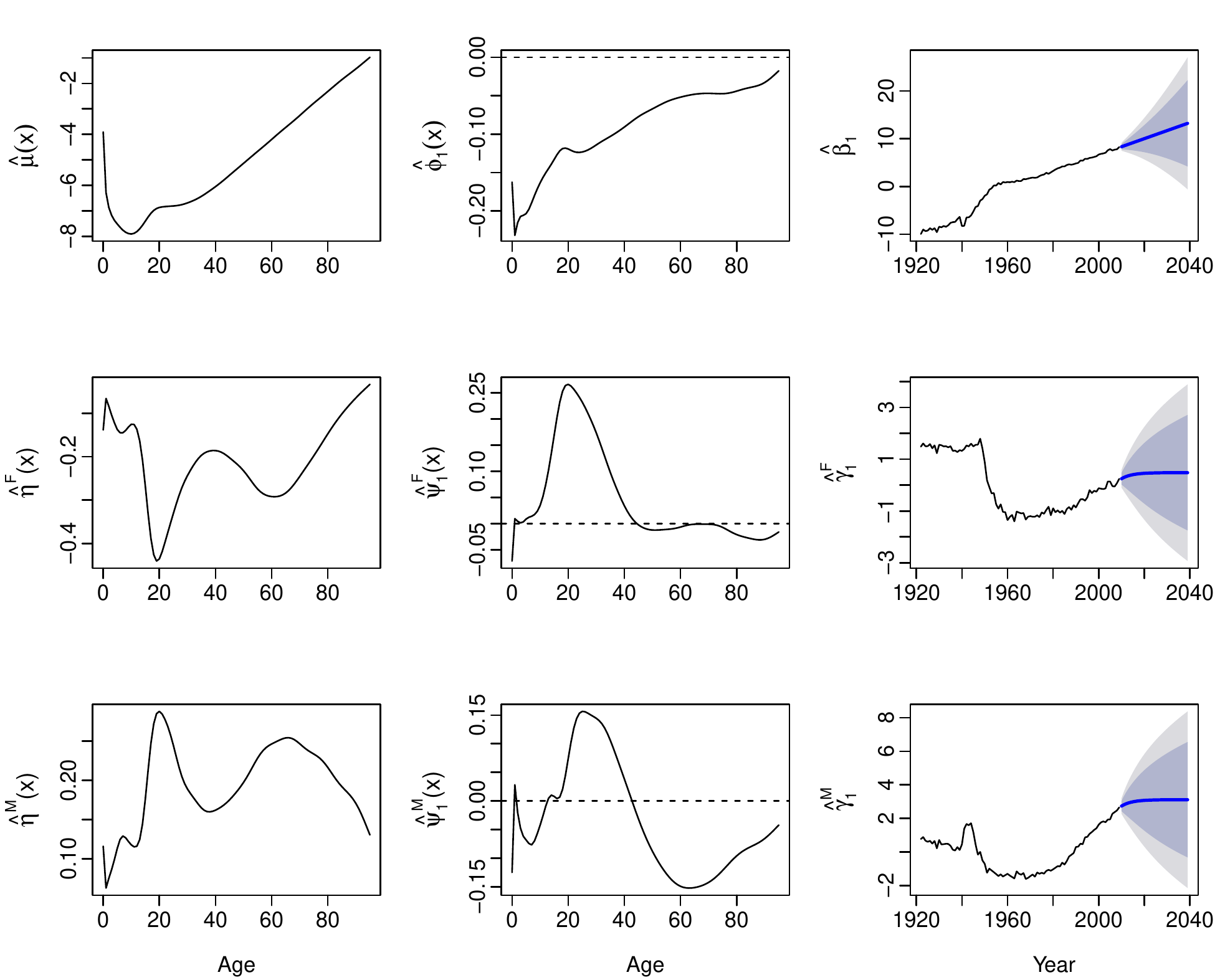}
  \caption{Estimated common mean function, first common functional principal component, and associated scores for the UK total mortality (top); estimated mean function deviation for females, first functional principal component, and associated scores for the UK female mortality (middle); estimated mean function deviation for males, first functional principal component, and associated scores for the UK male mortality (bottom). The dark and light gray regions show the 80\% and 95\% prediction intervals, respectively.}\label{Shang:fig:2}
\end{figure}

In the top panel of Fig.~\ref{Shang:fig:2}, we display the estimated common mean function $\widehat{\mu}(x)$, first estimated common functional principal component $\widehat{\phi}_1(x)$ and corresponding scores $\{\widehat{\beta}_{1,1},\widehat{\beta}_{2,1},\dots,\widehat{\beta}_{n,1}\}$ along with their 30-years-ahead forecasts. The first common functional principal component captures more than 98\% of the total variation in the age-specific total mortality. In the middle panel of Fig~\ref{Shang:fig:2}, we show the estimated mean function deviance of females from the overall mean function $\widehat{\eta}^{\text{F}}(x)$, first functional principal component for females $\widehat{\psi}_1^{\text{F}}(x)$ and corresponding scores $\{\widehat{\gamma}_{1,1}^{\text{F}},\widehat{\gamma}_{2,1}^{\text{F}},\dots,\widehat{\gamma}_{n,1}^{\text{F}}\}$ with 30-years-ahead forecasts. In the bottom panel of Fig.~\ref{Shang:fig:2}, we show the estimated mean function deviance of males from the overall mean function $\widehat{\eta}^{\text{M}}(x)$, first functional principal component for males $\widehat{\psi}_1^{\text{M}}(x)$ and corresponding scores $\{\widehat{\gamma}_{1,1}^{\text{M}},\widehat{\gamma}_{2,1}^{\text{M}},\dots,\widehat{\gamma}_{n,1}^{\text{M}}\}$ with 30-years-ahead forecasts. In this data set, the first three functional principal components explain at least 90\% of the remaining 10\% total variations for both females and males. Here, we display only the first functional principal component, which captures more than 64\% and 50\% of the remaining 10\% total variations for both females and males. Based on~\eqref{Shang:eq:2}, the proportion of variability explained by the total mortality is 94\% for females and 95\% for males.

\subsection{Forecast accuracy evaluation}

\subsubsection{Evaluation of point forecast accuracy}

We split our age- and sex-specific data into a training sample (including data from years 1 to $(n-30)$) and a testing sample (including data from years $(n-29)$ to $n$), where $n$ represents the total number of years in the data. Following the early work by \citet{HB08}, we implement an expanding window approach as it allows us to assess the forecast accuracy among methods for different forecast horizons. With the initial training sample, we produce one-to 30-year-ahead forecasts, and determine the forecast errors by comparing the forecasts with actual out-of-sample data. As the training sample increases by one year, we produce one-to 29-year-ahead forecasts and calculate the forecast errors. This process continues until the training sample covers all available data.

To measure the point forecast accuracy, we utilize the root mean squared forecast error (RMSFE), root maximum squared forecast error (Max RSFE), mean absolute forecast error (MAFE), maximum absolute forecast error (Max AFE) and mean forecast error (MFE). They are defined as:
\begin{align*}
\text{RMSFE}(h) &= \sqrt{\frac{1}{(31-h)\times p}\sum^n_{k=n-30+h}\sum^p_{i=1}\left[m_k(x_i) - \widehat{m}_k(x_i)\right]^2}, \\
\text{Max RSFE}(h) &= \sqrt{\max_{k,i}\left[m_k(x_i) - \widehat{m}_k(x_i)\right]^2}, \\
\text{MAFE}(h) &= \frac{1}{(31-h)\times p} \sum^n_{k=n-30+h}\sum^p_{i=1}\left|m_k(x_i) - \widehat{m}_k(x_i)\right|, \\
\text{Max AFE}(h) &= \max_{k,i} \left|m_k(x_i) - \widehat{m}_k(x_i)\right|, \\
\text{MFE}(h) &= \frac{1}{(31-h)\times p}\sum^n_{k=n-30+h}\sum^p_{i=1}\left[m_k(x_i) - \widehat{m}_k(x_i)\right],
\end{align*}
for $k=n-30+h,\dots,n$ and $h=1,\dots,30$, where $m_k(x_i)$ represents mortality rate at year $k$ in the forecasting period for age $x_i$, and $\widehat{m}_k(x_i)$ represents the point forecast.

\subsubsection{Evaluation of interval forecast accuracy}

To assess interval forecast accuracy, we use the interval score of \citet{GR07} \citep[see also][]{GK14}. For each year in the forecasting period, one-year-ahead to 30-year-ahead prediction intervals were calculated at the $(1-\alpha)\times 100\%$ prediction interval, with lower and upper bounds that are predictive quantiles at $\alpha/2$ and $1-\alpha/2$, denoted by $x_l$ and $x_u$. As defined by \citet{GR07}, a scoring rule for the interval forecast at age $x_i$ is
\begin{equation}
S_{\alpha}(x_l,x_u;x_i) =(x_u - x_l) + \frac{2}{\alpha}(x_l-x_i)I\{x_i<x_l\} + \frac{2}{\alpha}(x_i - x_u)I\{x_i>x_u\},
\end{equation}
where $I\{\cdot\}$ represents the binary indicator function, and $\alpha$ denotes the level of significance, customarily $\alpha=0.2$. A forecaster is rewarded for narrow prediction intervals, but incurs a penalty, the size of which depends on $\alpha$, if the observation misses the interval. The smallest interval score is the one that achieves the best tradeoff between empirical coverage probability and halfwidth of prediction interval.

For different ages and years in the forecasting period, the maximum and mean interval scores for each horizon are defined by
\begin{align*}
\max[S_{\alpha}(h)] &= \max_{k,i} S_{\alpha,k}(x_l,x_u;x_i), \\
\bar{S}_{\alpha}(h) &= \frac{1}{(31-h)\times p}\sum^n_{k=n-30+h}\sum^p_{i=1}S_{\alpha,k}(x_l,x_u;x_i),
\end{align*}
where $p$ represents the total number of ages or age groups in the evaluation data set. The best forecasting method is considered to be the one that produces the smallest maximum or mean interval score.

\subsection{Comparison of point forecast accuracy}

We compare the point forecast accuracy between the standard and robust multilevel functional data methods. As with the robust multilevel functional data method, it is necessary to specify a tuning parameter $\lambda$. When $\lambda\rightarrow \infty$, it corresponds to the standard multilevel functional data method, where no outlier can be detected. When $\lambda\rightarrow0$, it considers all observations as outliers. Here, we consider four different values for $\lambda=1.81, 2.33, 3, 3.29$, which reflects 90\%, 95\%, 98.3\% and 99\% of efficiency. For this data set, we found that the robust multilevel functional data method outperforms the standard multilevel functional data method. The optimal forecast accuracy is achieved when $\lambda = 1.81$, regardless which univariate time series forecasting method (rwf, ARIMA, ets) is used. Among the three univariate time series forecasting methods, the random walk with drift generally performs the best with the smallest forecast errors for female and male mortality rates and male life expectancy, whereas the ARIMA forecasting method produces the smallest forecast errors for female life expectancy.

{\small
\tabcolsep 0.29cm
\begin{longtable}{@{}llrrrrrrrrrr@{}}
\caption{Point forecast accuracy of age-specific mortality and life expectancy for females and males by different univariate time series forecasting methods, as measured by the Max AFE, Max RSFE, MAFE, RMSFE, and MFE. For mortality, the forecast errors were multiplied by 100 in order to keep two decimal places. The minimal forecast errors are highlighted in bold for females and males.}\\
\hline
& & \multicolumn{5}{c}{Mortality ($\times 100$)} &  \multicolumn{5}{c}{Life expectancy} \\
 Error & Sex & $1.81$ & $2.33$ & $3$  & $3.29$ & $\infty$ & $1.81$ & $2.33$ & $3$  & $3.29$ & $\infty$\\
\hline
  \endfirsthead
  \hline
& & \multicolumn{5}{c}{Mortality ($\times 100$)} &  \multicolumn{5}{c}{Life expectancy} \\
 Error & Sex & $1.81$ & $2.33$ & $3$  & $3.29$ & $\infty$ & $1.81$ & $2.33$ & $3$  & $3.29$ & $\infty$\\
\hline
\endhead
\hline \multicolumn{12}{r}{\emph{Continued on next page}}
\endfoot
\hline
\endlastfoot
 Max AFE 	& F & 6.11 & 6.18 & 6.25  & 6.29  & 7.18 & 2.23 & 2.27 & 2.33 & 2.34 & 3.24 \\ 
  (rwf) 	& M & \textBF{7.31} & 7.36 & 7.41 & 7.45 & 8.68 & \textBF{2.27} & 2.31 & 2.38 & 2.41 & 3.15 \\ 
  \\
  & F & \textBF{6.09} & 6.15 & 6.22 & 6.19 & 7.13 & \textBF{2.09} & 2.16 & 2.16 & 2.16 & 3.04  \\ 
  (ARIMA) & M & 7.90 & 7.93 & 8.06 & 8.10  & 8.55 & 2.88 & 2.88 & 2.97 & 2.98 & 3.23 \\ 
\\
& F & 6.50 & 6.58 & 6.62  & 6.64 & 7.60 & 2.74 & 2.79 & 2.80 & 2.80 & 3.65 \\ 
  (ets) & M & 7.99 & 8.01  & 8.09 & 8.07& 9.10 & 3.36 & 3.40  & 3.60  & 3.56 & 4.00  \\ 
\\
 Max RSFE & F  & \textBF{0.38} & 0.39 & 0.40 & 0.40 & 0.52 & 5.56 & 5.76 & 6.03 & 6.11 & 10.61 \\ 
  (rwf) & M  & \textBF{0.57} & 0.58 & 0.59 & 0.59 & 0.79 & \textBF{6.26} & 6.46 & 6.84 & 6.99 & 10.95 \\ 
  \\
  & F  & \textBF{0.38} & 0.39 & 0.40  & 0.39 & 0.51 & \textBF{4.83} & 5.13  & 5.17 & 5.16 & 9.29  \\ 
  (ARIMA) & M & 0.66 & 0.67 & 0.69 & 0.70  & 0.78 & 10.00 & 9.93 & 10.45 & 10.54   & 11.57  \\ 
 \\
  & F & 0.43 & 0.44 & 0.45 & 0.45 & 0.58 & 8.56 & 8.83 & 8.87 & 8.89 & 13.64  \\ 
  (ets) & M & 0.69 & 0.69 & 0.71 & 0.70 & 0.88 & 13.72 & 14.02 & 15.55 & 15.23 & 17.92  \\ 
\\
 MAFE & F & \textBF{0.42} & 0.43 & 0.46 & 0.48 & 0.68 & 1.59 & 1.63 & 1.73 & 1.78 & 2.53 \\ 
  (rwf) & M & \textBF{0.61} & 0.62  & 0.66 & 0.67& 0.82 & \textBF{1.76} & 1.80 & 1.90 & 1.94 & 2.47 \\ 
  \\
  & F & 0.45 & 0.46 & 0.48 & 0.48 & 0.64 & \textBF{1.48} & 1.54 & 1.60 & 1.64  & 2.22  \\ 
  (ARIMA) & M & 0.73 & 0.74 & 0.77 & 0.79 & 0.84 & 2.16 & 2.16 & 2.25  & 2.29 & 2.56 \\ 
 \\
  & F & 0.50 & 0.51 & 0.54 & 0.55 & 0.75 & 2.02 & 2.06 & 2.15 & 2.18 & 2.88  \\ 
  (ets) & M & 0.78 & 0.79 & 0.83 & 0.84 & 0.95 & 2.56 & 2.59 & 2.68 & 2.71 & 3.11 \\ 
\\
 RMSFE & F & \textBF{0.92} & 0.95  & 1.01 & 1.03 & 1.39  & 1.66 & 1.71 & 1.80 & 1.85 & 2.56 \\ 
  (rwf) & M  & \textBF{1.23} & 1.25 & 1.32 & 1.34 & 1.58 & \textBF{1.81} & 1.85 & 1.95 & 1.99 & 2.52 \\ 
  \\
 &  F & 0.99 & 1.02 & 1.04 & 1.05 & 1.34  & \textBF{1.54} & 1.61 & 1.66 & 1.70 & 2.27 \\ 
 (ARIMA) & M  & 1.44 & 1.44 & 1.51& 1.53 & 1.61  & 2.25 & 2.24 & 2.33 & 2.37 & 2.61  \\ 
 \\
  & F & 1.06 & 1.09 & 1.14 & 1.16 & 1.53 & 2.09 & 2.13 & 2.20 & 2.24 & 2.91 \\ 
  (ets) & M & 1.50 & 1.53  & 1.59 & 1.61 & 1.81 & 2.63 & 2.67 & 2.75 &  2.78 & 3.16 \\ 
\\
 MFE & F & \textBF{-0.33} & -0.35 & -0.39  & -0.41 & -0.67 & 1.58 & 1.62 & 1.72  & 1.77 & 2.53 \\ 
  (rwf) & M  & \textBF{-0.36} & -0.38  & -0.45 & -0.47 & -0.78 & \textBF{1.71} & 1.75 & 1.86  & 1.91 & 2.47 \\ 
  \\
  & F & -0.37 & -0.39 & -0.41 & -0.42  & -0.63 & \textBF{1.47} & 1.53 & 1.59  & 1.63 & 2.22 \\ 
  (ARIMA) & M & -0.52 & -0.52  & -0.58 & -0.61 & -0.80 & 2.09 & 2.10 & 2.20  & 2.24 & 2.56 \\ 
  \\
  & F & -0.44 & -0.45 & -0.48 & -0.50 & -0.74 & 2.02 & 2.05 & 2.14  & 2.18 & 2.88 \\ 
  (ets) & M & -0.60 & -0.62 & -0.67 & -0.69 & -0.92 & 2.51 & 2.54 & 2.65 & 2.68 & 3.11 \\ 
\end{longtable}}

\subsection{Comparison of interval forecast accuracy}

The prediction intervals for age-specific mortality are obtained from~\eqref{Shang:eq:multilevel_interval}, whereas the prediction intervals for life expectancy are obtained from the percentiles of simulated life expectancies obtained from simulated forecast mortality rates as described by \citet{HB08}. Based on the  mean interval scores in Table~\ref{Shang:tab:2}, we found the robust multilevel functional data method outperforms the standard multilevel functional data method. The ARIMA forecasting method gives the smallest interval scores for females when $\lambda=2.33$, whereas the exponential smoothing method performs the best for males when $\lambda=1.81$. 

\begin{table}[!t]
\centering
\caption{Interval forecast accuracy of mortality and life expectancy for females and males by different univariate time series forecasting methods, as measured by maximum interval score and mean interval score. For mortality, the interval scores were multiplied by 100 in order to keep two decimal places. The minimal interval scores are highlighted in bold for females and males.}\label{Shang:tab:2}
\tabcolsep 0.15cm
\begin{tabular}{@{}llrrrrrrrrrr@{}}\toprule
	& 	& 	 \multicolumn{5}{c}{Mortality ($\times 100$)} & \multicolumn{5}{c}{Life expectancy} \\
Error & Sex & 1.81 & 2.33 & 3 & 3.29 & $\infty$ & 1.81 & 2.33 & 3 & 3.29 & $\infty$ \\\midrule 
Max interval score & F & 26.17 & 25.72 & 27.64 & 27.93 & 33.40 & 11.84 & 12.23 & 13.09 & 13.09 & 18.26 \\
  (rwf) & M & \textBF{25.12} & 26.72 & 28.73 & 29.77 & 48.94 & 9.86 & 10.16 & 11.14 & 11.18 & 16.66 \\ 
  \\
  & F & 17.20 & 16.13 & 17.30 & 17.34 & 18.89 & 6.13 & \textBF{5.92} & 7.74 & 7.60 & 6.52 \\ 
  (ARIMA) & M & 33.39 & 33.31 & 33.39 & 33.75 & 38.01 & 15.37 & 15.04 & 16.86 & 16.71 & 12.04 \\ 
\\
  & F & 16.06 & 15.57 & 15.50 & \textBF{15.07} & 18.24 & 6.72 & 6.79 & 6.76 & 6.42 & 7.25 \\ 
  (ets) & M & 26.39 & 27.40 & 26.73 & 26.40 & 42.81 & \textBF{8.35} & 9.25 & 9.25 & 9.84 & 12.96 \\ 
  \\
  \\
  Mean  interval score & F & 2.27 & 2.36 & 2.57 & 2.69 & 3.42 & 8.04 & 8.47 & 9.25 & 9.66 & 13.50 \\ 
  (rwf) & M & 3.11 & 3.23 & 3.49 & 3.62 & 4.38 & 8.18 & 8.57 & 9.21 & 9.56 & 12.35 \\ 
 \\
  & F & 1.49 & \textBF{1.48} & 1.57 & 1.62 & 1.52 & 4.75 & \textBF{4.67} & 5.09 & 5.36 & 4.97 \\ 
  (ARIMA) & M & 3.68 & 3.50 & 3.74 & 3.62 & 3.11 & 11.52 & 10.76 & 11.88 & 11.13 & 9.40 \\ 
 \\
  & F & 1.56 & 1.58 & 1.56 & 1.51 & 1.48 & 5.95 & 5.84 & 5.83 & 5.50 & 5.35 \\ 
  (ets) & M & \textBF{2.91} & 3.01 & 2.98 & 3.08 & 3.30 & \textBF{7.38} & 7.88 & 7.82 & 8.08 & 8.54 \\ 
\bottomrule	
\end{tabular}
\end{table}

\section{Conclusion}\label{Shang:sec:4}

In this paper, we put forward a robust multilevel functional data method to forecast age-specific mortality and life expectancy at birth for a group of populations. This method inherits the smoothness property a functional time series possesses, thus missing data can be naturally dealt with. In addition, this method is a robust approach that can handle the presence of outliers. 

As demonstrated by the empirical studies consisting of two sub-populations in the UK, we found that the robust multilevel functional data method produces more accurate forecasts than the standard multilevel functional data method in the presence of outlying years largely due to World Wars and Spanish flu pandemic in the UK. Based on the averaged forecast errors, the robust multilevel functional data method with $\lambda=1.81$ gives the most accurate point forecasts among all we considered. Furthermore, we consider three univariate time series forecasting methods and compare their point and interval forecast accuracy. Among the three univariate time series forecasting methods, the random walk with drift generally performs the best for female and male mortality rates and male life expectancy, whereas the ARIMA forecasting method produces the smallest forecast errors for female life expectancy. Based on the mean interval scores, the ARIMA forecasting method gives the smallest interval scores for females when $\lambda=2.33$, whereas the exponential smoothing method performs the best for males when $\lambda=1.81$. It is a straightforward extension to average forecasts obtained from all three univariate time series forecasting methods in hope to improve forecast accuracy. Although $\lambda=1.81$ works well in the data set considered, the optimal selection of $\lambda$ remains as a challenge and an open problem for future research. 

Another research topics are that although the proposed methods are demonstrated using the UK data, the methodology can easily be extended to mortality data from other countries. Furthermore, the multilevel functional data model captures correlation between a group of populations based on sex, but the methodology can also be extended to some other characteristics, such as state or ethnic group. It would also be interesting to investigate the performance of this robust multilevel functional data method for various lengths of functional time series.

\newpage
\bibliographystyle{apalike}
\bibliography{coherent}

\begin{thebibliography}{}

\bibitem[Alkema et~al., 2011]{ARG+11}
Alkema, L., Raftery, A.~E., Gerland, P., Clark, S.~J., Pelletier, F., Buettner,
  T., and Heilig, G.~K. (2011).
\newblock Probabilistic projections of the total fertility rate for all
  countries.
\newblock {\em Demography}, 48(3):815--839.

\bibitem[Booth, 2006]{Booth06}
Booth, H. (2006).
\newblock Demographic forecasting: 1980-2005 in review.
\newblock {\em International Journal of Forecasting}, 22(3):547--581.

\bibitem[Booth and Tickle, 2008]{BT08}
Booth, H. and Tickle, L. (2008).
\newblock Mortality modelling and forecasting: A review of methods.
\newblock {\em Annals of Actuarial Science}, 3(1-2):3--43.

\bibitem[Chiou, 2012]{Chiou12}
Chiou, J.-M. (2012).
\newblock Dynamical functional prediction and classification, with application
  to traffic flow prediction.
\newblock {\em Annals of Applied Statistics}, 6(4):1588--1614.

\bibitem[Crainiceanu and Goldsmith, 2010]{CG10}
Crainiceanu, C.~M. and Goldsmith, J.~A. (2010).
\newblock {Bayesian functional data analysis using WinBUGS}.
\newblock {\em Journal of Statistical Software}, 32(11).

\bibitem[Crainiceanu et~al., 2009]{CSD09}
Crainiceanu, C.~M., Staicu, A.-M., and Di, C.-Z. (2009).
\newblock {Generalized multilevel functional regression}.
\newblock {\em Journal of the American Statistical Association},
  104(488):1550--1561.

\bibitem[Delwarde et~al., 2006]{DDG+06}
Delwarde, A., Denuit, M., {Guill\'{e}n}, M., and {Vidiella-i-Anguera}, A.
  (2006).
\newblock {Application of the Poisson log-bilinear projection model to the G5
  mortality experience}.
\newblock {\em Belgian Actuarial Bulletin}, 6(1):54--68.

\bibitem[Di et~al., 2009]{DCC+09}
Di, C.-Z., Crainiceanu, C.~M., Caffo, B.~S., and Punjabi, N.~M. (2009).
\newblock Multilevel functional principal component analysis.
\newblock {\em Annals of Applied Statistics}, 3(1):458--488.

\bibitem[Girosi and King, 2008]{GK08}
Girosi, F. and King, G. (2008).
\newblock {\em {Demographic Forecasting}}.
\newblock Princeton University Press, Princeton.

\bibitem[Gneiting and Katzfuss, 2014]{GK14}
Gneiting, T. and Katzfuss, M. (2014).
\newblock Probabilistic forecasting.
\newblock {\em Annual Review of Statistics and Its Applications}, 1:125--151.

\bibitem[Gneiting and Raftery, 2007]{GR07}
Gneiting, T. and Raftery, A.~E. (2007).
\newblock Strictly proper scoring rules, prediction and estimation.
\newblock {\em Journal of the American Statistical Association},
  102(477):359--378.

\bibitem[Greven et~al., 2010]{GCC+10}
Greven, S., Crainiceanu, C., Caffo, B., and Reich, D. (2010).
\newblock Longitudinal functional principal component analysis.
\newblock {\em Electronic Journal of Statistics}, 4:1022--1054.

\bibitem[He and Ng, 1999]{HN99}
He, X. and Ng, P. (1999).
\newblock {COBS: Qualitatively constrained smoothing via linear programming}.
\newblock {\em Computational Statistics}, 14:315--337.

\bibitem[Hubert et~al., 2005]{HRB05}
Hubert, M., Rousseeuw, P., and Branden, K. (2005).
\newblock {ROBPCA: A new approach to robust principal component analysis}.
\newblock {\em Technometrics}, 47(1):64--79.

\bibitem[Hubert et~al., 2002]{HRV02}
Hubert, M., Rousseeuw, P., and Verboven, S. (2002).
\newblock A fast method for robust principal components with applications to
  chemometrics.
\newblock {\em Chemometrics and Intelligent Laboratory Systems},
  60(1-2):101--111.

\bibitem[{Human Mortality Database}, 2015]{HMD13}
{Human Mortality Database} (2015).
\newblock {\em {University of California, Berkeley (USA), and Max Planck
  Institute for Demographic Research (Germany)}}.
\newblock Accessed at 8 March 2013. URL: \url{http://www.mortality.org}.

\bibitem[Hyndman and Booth, 2008]{HB08}
Hyndman, R.~J. and Booth, H. (2008).
\newblock {Stochastic population forecasts using functional data models for
  mortality, fertility and migration}.
\newblock {\em International Journal of Forecasting}, 24(3):323--342.

\bibitem[Hyndman and Ullah, 2007]{HU07}
Hyndman, R.~J. and Ullah, M.~S. (2007).
\newblock {Robust forecasting of mortality and fertility rates: A functional
  data approach}.
\newblock {\em Computational Statistics \& Data Analysis}, 51(10):4942--4956.

\bibitem[Lee, 2006]{Lee06}
Lee, R.~D. (2006).
\newblock Mortality forecasts and linear life expectancy trends.
\newblock In Bengtsson, T., editor, {\em Perspectives on mortality forecasting.
  Vol. III. The linear rise in life expectancy: History and prospects},
  number~3 in Social Insurance Studies, pages 19--39. {Swedish National Social
  Insurance Board}, Stockholm.

\bibitem[Lee and Carter, 1992]{LC92}
Lee, R.~D. and Carter, L.~R. (1992).
\newblock {Modeling and forecasting U.S. mortality}.
\newblock {\em Journal of the American Statistical Association},
  87(419):659--671.

\bibitem[Li, 2013]{Li13}
Li, J. (2013).
\newblock {A Poisson common factor model for projecting mortality and life
  expectancy jointly for females and males}.
\newblock {\em Population Studies}, 67(1):111--126.

\bibitem[Li and Lee, 2005]{LL05}
Li, N. and Lee, R. (2005).
\newblock {Coherent mortality forecasts for a group of population: An extension
  of the Lee-Carter method}.
\newblock {\em Demography}, 42(3):575--594.

\bibitem[Li et~al., 2013]{LLG13}
Li, N., Lee, R., and Gerland, P. (2013).
\newblock {Extending the Lee-Carter method to model the rotation of age
  patterns of mortality decline for long-term projections}.
\newblock {\em Demography}, 50(6):2037--2051.

\bibitem[Preston et~al., 2001]{PHG01}
Preston, S.~H., Heuveline, P., and Guillot, M. (2001).
\newblock {\em {Demography: Measuring and Modelling Population Process}}.
\newblock Blackwell, Oxford, UK.

\bibitem[Raftery et~al., 2013]{RCG+13}
Raftery, A.~E., Chunn, J.~L., Gerland, P., and \v{S}ev\v{c}\'{i}kov\'{a}, H.
  (2013).
\newblock Bayesian probabilistic projections of life expectancy for all
  countries.
\newblock {\em Demography}, 50(3):777--801.

\bibitem[Raftery et~al., 2014]{RLG14}
Raftery, A.~E., Lalic, N., and Gerland, P. (2014).
\newblock Joint probabilistic projection of female and male life expectancy.
\newblock {\em Demographic Research}, 30:795--822.

\bibitem[Raftery et~al., 2012]{RLS+12}
Raftery, A.~E., Li, N., \v{S}ev\v{c}\'{i}kov\'{a}, H., Gerland, P., and Heilig,
  G.~K. (2012).
\newblock {Bayesian probabilistic population projection for all countries}.
\newblock {\em Proceedings of the National Academy of Sciences of the United
  States of America}, 109(35):13915--13921.

\bibitem[Renshaw and Haberman, 2003]{RH03}
Renshaw, A.~E. and Haberman, S. (2003).
\newblock {Lee-Carter mortality forecasting with age-specific enhancement}.
\newblock {\em Insurance: Mathematics and Economics}, 33(2):255--272.

\bibitem[Rice and Silverman, 1991]{RS91}
Rice, J. and Silverman, B. (1991).
\newblock Estimating the mean and covariance structure nonparametrically when
  the data are curves.
\newblock {\em Journal of the Royal Statistical Society. Series B},
  53(1):233--243.

\bibitem[Shang, 2016]{Shang16}
Shang, H.~L. (2016).
\newblock {Mortality and life expectancy forecasting for a group of populations
  in developed countries: A multilevel functional data method}.
\newblock {\em Annals of Applied Statistics}, in press.

\bibitem[Shang et~al., 2011]{SBH11}
Shang, H.~L., Booth, H., and Hyndman, R.~J. (2011).
\newblock Point and interval forecasts of mortality rates and life expectancy:
  A comparison of ten principal component methods.
\newblock {\em Demographic Research}, 25(5):173--214.

\bibitem[Tickle and Booth, 2014]{BT14}
Tickle, L. and Booth, H. (2014).
\newblock {The longevity prospects of Australian seniors: An evaluation of
  forecast method and outcome}.
\newblock {\em Asia-Pacific Journal of Risk and Insurance}, 8(2):259--292.

\bibitem[\v{S}ev\v{c}ikov\'{a} et~al., 2015]{SLK+15}
\v{S}ev\v{c}ikov\'{a}, H., Li, N., Kantorov\'{a}, V., Gerland, P., and Raftery,
  A.~E. (2015).
\newblock {Age-specific mortality and fertility rates for probabilistic
  population projections}.
\newblock In Schoen, R., editor, {\em Dynamic Demographic Analysis}, volume~39
  of {\em The Springer Series on Demographic Methods and Population Analysis},
  pages 285--310. Springer.

\bibitem[Wi\'{s}niowski et~al., 2015]{WSB+15}
Wi\'{s}niowski, A., Smith, P. W.~F., Bijak, J., Raymer, J., and Forster, J.~J.
  (2015).
\newblock {Bayesian population forecasting: Extending the Lee-Carter method}.
\newblock {\em Demography}, 52(3):1035--1059.

\bibitem[Yao et~al., 2005]{YMW05}
Yao, F., M\"{u}ller, H.-G., and Wang, J. (2005).
\newblock Functional data analysis for sparse longitudinal data.
\newblock {\em Journal of the American Statistical Association},
  100(470):577--590.

\end{thebibliography}

\end{document}